\documentstyle[11pt]{article}

\newlength{\dinwidth}
\newlength{\dinmargin}
\begin{document}
\def\Db{\overline{D}}
\def\nle{{\stackrel{<}{\sim}}}
\def\nge{{\stackrel{>}{\sim}}}
\def\st{\widetilde{t}}
\def\stl{\st_{1}}
\def\sth{\st_{2}}
\def\mstl{m_{\stl}}
\def\mt{m_{t}}
\def\mb{m_{b}}
\def\mz{m_{Z}}
\def\mw{m_{W}}
\def\tht{\theta_{t}}
\def\tew{\theta_{W}}
\def\gev{{\rm GeV}}
\def\slepton{\widetilde \ell}
\def\squark{\widetilde q}
\def\sdr{{\widetilde d}_{R}}
\def\msq{m_{\squark}}
\def\msl{m_{\slepton}}
\def\photino{\widetilde \gamma}
\def\sfermion{\widetilde f}
\def\sf{{\widetilde f}}
\def\sneutrino{\widetilde \nu}
\def\selectron{\widetilde e}
\def\se{{\widetilde e}}
\def\sd{{\widetilde d}}
\def\mse{m_{\selectron}}
\def\msel{m_{\selectron_{L}}}
\def\msdr{m_{\sdr}}
\def\msn{m_{\sneutrino}}
\def\gluino{\widetilde g}
\def\ddf{{\rm d}}
\def\lam{\lambda'_{131}}
\def\zino{{\widetilde{Z}}}
\def\wino{{\widetilde{W}}}
\def\cbar{\overline{c}}
\def\sz1{{\widetilde{Z}}_{1}}
\def\szi{{\widetilde{Z}}_{i}}
\def\swl{{\widetilde{W}}_{1}}
\def\swk{{\widetilde{W}}_{k}}
\def\msz1{m_{\sz1}}
\def\mswl{m_{\swl}}
\def\mszi{m_{\szi}}
\def\mswk{m_{\swk}}
\def\ino{\widetilde{\chi}}
\def\Rb{\slash\hspace{-5pt}R}
\def\misE{\slash\hspace{-8pt}E}
\def\misEt{\slash\hspace{-8pt}E_{T}}
\def\misP{\slash\hspace{-7pt}P_{T}}
\def\tanbe{\tan\beta}
\def\ccbar{\overline{c}}
\def\lam{\lambda'_{131}}
\def\lams{\lambda^{'2}_{131}}
\def\frb{F_{\rm RB}}
\def\i{{\rm i}}
\def\xq{(x, Q^{2})}
\def\qz{{{Q^2}\over{Q^2 +\mz^2}}}
\def\utg{(Q^2-sx-\mstl^{2})}
\def\sh{\hat{s}}
\def\th{\hat{t}}
\def\uh{\hat{u}}
\def\stg{(\sh-\mstl^{2})^{2}+\mstl^{2}\Gamma_{\stl}^{2}}
\def\goto{\rightarrow}

\begin{flushright}
ITP-SU-94/01 \\
March, 1994
\end{flushright}
\begin{center}
\renewcommand{\thefootnote}{\fnsymbol{footnote}}
\vglue 0.6cm
 {\bf        \vglue 10pt
Signatures of Scalar Top with R-parity Breaking Coupling at HERA
\\}
\vglue 1.0cm
{Tadashi Kon \\}
\baselineskip=13pt
{\it Faculty of Engineering, Seikei University\\}
\baselineskip=12pt
{\it Tokyo, 180, Japan\\}
\vglue 0.3cm
{\tenrm Tetsuro Kobayashi \\}
\baselineskip=13pt
{\tenit Department of Physics, Tokyo Metropolitan University\\}
\baselineskip=12pt
{\tenit Tokyo, 192-03, Japan
\footnote{After 1 April, 1994,
{\it Fukui University of Engineering,
Fukui, 910, Japan}}
\\}
\vglue 0.3cm
{\tenrm and\\}
\vglue 0.3cm
{\tenrm Shoichi Kitamura \\}
\baselineskip=13pt
{\tenit Department of Physics, Tokyo Metropolitan University\\}
\baselineskip=12pt
{\tenit Tokyo, 192-03, Japan}
\vglue 0.8cm
{ABSTRACT}
\end{center}
\vglue -0.3cm
 \noindent
\begin{quotation}
In the framework of the minimal supersymmetric standard model
with an R-parity breaking coupling of the scalar top quark (stop)
we investigate production processes
and decay properties of the stop at HERA energies.
The model is characterized by a light stop possibly
lighter than the other squarks.
We show that the stop could be singly produced not only
in the neutral current processes
but also in associated processes whose final states contain
some heavy flavor quarks, bottom and top quarks.
These signatures would be useful to discriminate the stop from
leptoquarks.

\end{quotation}
\vglue 0.6cm
\baselineskip=14pt


  The search for the supersymmetric particles (sparticles) with masses
below an order of the magnitude of 1TeV is one of the most important
purpose of the present collider experiments.
For this purpose theoretical physicists should give answers to
following questions ;
{\romannumeral 1}) "which sparticle will be discovered
first?" and
{\romannumeral 2}) "what signature will be expected in such sparticle
production?"
Answers are obviously model dependent.
The simplest model is the minimal supersymmetric standard model
(MSSM) with conserved R-parity \cite{Haber} ;
\begin{equation}
R\equiv(-)^{2S+3B+L},
\end{equation}
where $S$, $B$ and $L$ denote the spin, baryon and lepton numbers,
respectively.
In the model the sparticle expected to be discovered first, i.e.,
the lightest charged sparticle, is the slepton $\slepton$,
the lighter chargino $\swl$ or the stop $\stl$.
Irrespective of kinds of the lightest charged sparticles,
there is a distinctive signature from the sparticle production,
the large missing energies $\misE$
carried off by the lightest sparticle (LSP).
The neutralness and the R-parity conservation guarantee
the stablility and the very weak interaction with the matter
in the detector.

  Besides the MSSM with the conserved R-parity,
there are models with the R-parity breaking (RB) couplings in the
superpotential \cite{BARGER},
\begin{equation}
W_{\Rb}=\lambda_{ijk}L_{i}L_{j}\overline{E}_{k}+
   \lambda'_{ijk}L_{i}Q_{j}\overline{D}_{k}+
   \lambda''_{ijk}\overline{U}_{i}\overline{U}_{j}\overline{D}_{k},
\label{RBc}
\end{equation}
where $L$, $E$, $Q$, $U$ and $D$ denote the appropriate chiral
supermultiplets and $i\sim k$ are generation indices.
We should note that these couplings are not forbidden by
the gauge symmetry as well as by the supersymmetry.
The first two terms violate the lepton number $L$ and the last term
violates the baryon number $B$.
Consequently, the RB couplings in the
supersymmetric models may be required
in order to explain the cosmic baryon number violation, the origin of
the masses and the magnetic moments of neutrinos and some
interesting rare processes in terms of the $L$ and/or $B$ violation.
Here we should keep in mind that we have to take the RB couplings
as $\lambda$, $\lambda'$ $\ll$ 1 or $\lambda''$ $\ll$ 1 to guarantee
the stability against the fast proton decay.
If we take sizable RB couplings the LSP can decay into the ordinary
particles.
In this case the typical signatures of the sparticle production would
be multi-jets and/or multi-leptons instead of the large $\misE$.

  Here we focus our attention on the
stop $\stl$ and investigate its production
mechanisms and decay processes realized at HERA
in the framework of the MSSM with the RB couplings of the stop.
The RB couplings
$W=\lambda'_{1jk}L_{1}Q_{j}{\bar{D}}_{k}$
originated from the second term of the RB superpotential (\ref{RBc})
are the most suitable for the $ep$ collider experiments
at HERA because the squarks will be produced through the $s$-channel
in the $e$-$q$ sub-processes.
In the MSSM the stop mass could be lower than that of the
other squarks in a model \cite{HK} because of the
high expected mass of the top quark.
The expected mass of the stop could be within the reach of HERA.
The production of squarks with RB couplings
in the first and second generation
at HERA has been discussed extensively in Ref.~\cite{susana}.

In previous papers \cite{stoprb,proc,Erice,full}
we have considered
the stop $\stl$ production through the $s$-channel in
neutral current (NC) processes :
\begin{equation}
ep \rightarrow (\stl X) \rightarrow eqX,
\label{RBNC}
\end{equation}
and we have shown that we could get a clear signal as a sharp peak
in the Bjorken variable $x$ distribution.
However, one of the leptoquarks ${\widetilde{S}}_{1/2}$
\cite{leptoquark}
with the charge $Q=-2/3$ would give the same signature as that of
the RB stop,
if the stop has BR($\stl\rightarrow ed$)$\simeq$100\%.
Note that this situation corresponds to the case of
$\mstl$$<$$m_{t}+m_{\sz1}$, $m_{b}+m_{\swl}$.
In this case it is difficult to discriminate the stop from the
leptoquark ${\widetilde{S}}_{1/2}$.
In this paper we generalize our calculation including the case of
$\mstl$$>$$m_{t}+m_{\szi}$ or $\mstl$$>$$m_{b}+m_{\swk}$
because HERA could search the heavy stop with mass
$\mstl$ $\nle$ 300GeV.
We search for a possible experimental
observable in the RB stop production in $ep$ collisions.

In the framework of the MSSM,
scalar fermion mass matrices
in the ($\sf_{L}$, $\sf_{R}$) basis are expressed by
\begin{equation}\renewcommand{\arraystretch}{1.3}
{\cal M}^{2}_{\sfermion}=\left(
                 \begin{array}{cc}
                   m^{2}_{\sfermion_{L}} & a_{f}m_{f} \\
                   a_{f}m_{f} & m^{2}_{\sfermion_{R}}
                 \end{array}
                \right),
\label{matrix}
\end{equation}
where $m_{\sfermion_{L, R}}$ and $a_{f}$ are the SUSY mass parameters
and $m_{f}$ denote the ordinary fermion masses.
We can see from Eq.~(\ref{matrix}) that for the sleptons and
the squarks
except for the stops, the left and right handed sfermions are mass
eigenstates in good approximation owing to small fermion masses
in the off-diagonal elements of the mass matrices.
On the other hand, the large mixing between the left and right
handed stops will be expected because of the large top-quark mass
\cite{HK},
and the mass eigenstates are expressed by
\begin{equation}
\left({\stl\atop\sth}\right)=
\left(
{\st_{L}\,\cos\tht-\st_{R}\,\sin\tht}
\atop
{\st_{L}\,\sin\tht+\st_{R}\,\cos\tht}
\right),
\end{equation}
where $\tht$ denotes the mixing angle of stops :
\begin{eqnarray}
\sin 2\tht=
{\frac{2a_{t}\,m_{t}}
  {\sqrt{(m^{2}_{\st_{L}}-m^{2}_{\st_{R}})^{2}
          +4a^{2}_{t}\,m^{2}_{t}}}}, \\
\cos 2\tht=
{\frac{m^{2}_{\st_{L}}-m^{2}_{\st_{R}}}
  {\sqrt{(m^{2}_{\st_{L}}-m^{2}_{\st_{R}})^{2}
          +4a^{2}_{t}\,m^{2}_{t}}}}.
\label{sintht}
\end{eqnarray}
We can easily calculate the mass eigenvalues of the stops as
\begin{equation}
m^{2}_{\stl\atop\sth}
         ={\frac{1}{2}}\left[ m^{2}_{\st_{L}}+m^{2}_{\st_{R}}
             \mp \left( (m^{2}_{\st_{L}}-m^{2}_{\st_{R}})^{2}
            +(2a_{t}m_{t})^{2}\right)^{1/2}\right].
\label{stopmass}
\end{equation}
We find that if SUSY mass parameters and the top
mass are the same order of magnitude, the cancellation could
occur in the expression for the
lighter stop mass Eq.~(\ref{stopmass}).
Moreover,
the diagonal mass parameters $m^{2}_{\st_{L}}$ and $m^{2}_{\st_{R}}$
in Eq.(\ref{matrix}) have possibly small values
owing to the large negative contributions proportional
to the top quark Yukawa coupling which is determined by
the renormalization group equations in the mininal
supergravity GUTs \cite{RGE}.
So we get one light stop $\stl$ lighter than
the first and the second generation squarks for a wide range of
the SUSY parameters.
Note that $\stl$ could be even lighter than the top quark.
After the mass diagonalization
we can obtain the interaction Lagrangian in terms of the
mass eigenstate $\stl$.
In particular the relevant RB coupling of the stop is
obtained by
\begin{equation}
{\cal L}_{\rm int}=
{\lam}\cos\theta_{t}(\st_{1}\bar{d}P_{L}e
                         +\st_{1}^{*}\bar{e}P_{R}d),
\label{rbL}
\end{equation}
which is originated from the second term of the RB superpotential
(\ref{RBc}).
The coupling (\ref{rbL})
is the most suitable for the $ep$ collider experiments
at HERA because the stop will be produced through the $s$-channel
in the $e$-$d$ sub-processes.
For simplicity we take $\lam$ to be only non-zero coupling parameter
in the following.
The upper bound on the strength of copling $\lam$
has been investigated through
the low energy experiments \cite{BARGER} and the neutrino physics
\cite{neutrino}.
The most stringest present bound, $\lam$ $\nle$ 0.3,
comes from the atomic parity violation experiments \cite{BARGER}.

Actually, the stop can decay into the various final states :
\begin{eqnarray*}
\stl &\to& t\,\szi   \qquad\qquad\qquad\qquad\qquad\qquad({\rm a}) \\
 &\to& b\,\swk   \ \ \quad\qquad\qquad\qquad\qquad\qquad({\rm b})\\
 &\to& b\,\ell\,\sneutrino \qquad\qquad\qquad\qquad\qquad\qquad({\rm c})\\
 &\to& b\,\nu\,\slepton \qquad\qquad\qquad\qquad\qquad\qquad({\rm d})\\
 &\to& b\,W\,\szi \ \qquad\qquad\qquad\qquad\qquad\qquad({\rm e})\\
 &\to& b\,f\,\overline{f}\,\szi \qquad\qquad\qquad\qquad\qquad
\qquad({\rm f})\\
 &\to& c\,\sz1, \quad\qquad\qquad\qquad\qquad\qquad\qquad({\rm g})\\
 &\to& e\,d, \ \ \quad\qquad\qquad\qquad\qquad\qquad\qquad({\rm h})
\end{eqnarray*}
where $\szi$ ($i=1\sim4$), $\swk$ ($k=1, 2$), $\sneutrino$
and $\slepton$, respectively, denote
the neutralino, the chargino, the sneutrino and the
slepton.
(a) $\sim$ (g) are the R-parity conserving decay modes, while
(h) is realized by the RB coupling (\ref{rbL}).
If we consider the RB coupling $\lam$ $>$ 0.01,
which corresponds to the coupling strength detectable at HERA,
the decay modes (c) to (g) are negligible
due to their large power of $\alpha$
arising from multiparticle finel state or one loop contribution.
So there left the two body modes (a), (b) and (h).
The formulae of the decay widths for
each mode are respectively given by
\begin{eqnarray}
\Gamma(\stl \to t\,\szi)&&=
{\frac{\alpha}{2\mstl^{3}}}\lambda^{1/2}(\mstl^2, \mt^2, \mszi^2)
\nonumber \\ && \times
\left[\left(|F_L|^2+|F_R|^2\right)\left(\mstl^2-\mt^2-\mszi^2\right)
-4\mt\mszi{\rm Re}\left(F_R F_L^*\right)\right], \\
 && F_L\equiv{\frac{\mt N'^*_{i4}\cos\tht}{2\mw\sin\tew\sin\beta}}
+e_u\left(N'^*_{i1}-\tan\tew N'^*_{i2}\right)\sin\tht , \\
 && F_R\equiv
\left(e_u N'_{i1}+{\frac{1/2-e_u\sin^2\tew}{\cos\tew\sin\tew}}
N'_{i2}\right)\cos\tht -
{\frac{\mt N'_{i4}\sin\tht}{2\mw\sin\tew\sin\beta}},
\end{eqnarray}
\begin{eqnarray}
\Gamma(\stl \to b\,\swk)&&=
{\frac{\alpha}{4\sin^2\tew\mstl^{3}}}
\lambda^{1/2}(\mstl^2, \mb^2, \mswk^2)
\nonumber \\ &&  \times
\left[\left(|G_L|^2+|G_R|^2\right)\left(\mstl^2-\mb^2-\mswk^2\right)
-4\mb\mswk{\rm Re}\left(G_R G_L^*\right)\right], \\
 && G_L\equiv-{\frac{\mb U^*_{k2}\cos\tht}{\sqrt{2}\mw\cos\beta}} , \\
 && G_R\equiv V_{k1}\cos\tht+
{\frac{\mt V_{k2}\sin\tht}{\sqrt{2}\mw\sin\beta}}
\end{eqnarray}
and
\begin{equation}
\Gamma(\stl \to e\,d)={\frac{\lambda_{131}^{'2}}{16\pi}}\cos^2\tht\mstl,
\end{equation}
where $\lambda(x,y,z)$ $\equiv$ $x^2+y^2+z^2-2xy-2yz-2zx$.
$N'_{ij}$ and, $V_{kl}$ and $U_{kl}$ respectively
read the neutralino and chargino mixing angles \cite{Haber,GH}.
The mixing angles as well as the masses $\mszi$ and $\mswk$ are
determined by the basic parameters in the MSSM
($\mu$, $\tanbe$, $M_2$), where
$M_2$, $\tanbe$ and $\mu$ denote the soft breaking mass for SU(2) gaugino,
the ratio of two Higgs vacuum expectation values ($=v_{2}/v_{1}$)
and the supersymmetric Higgs mass parameter, respectively.
After all, we have input parameters
($\mu$, $\tanbe$, $M_2$, $\mstl$, $\tht$, $\mt$, $\lam$)
needed to calculate the decay widths and the
branching ratio of the stop.

If the stop is heavy enough with mass
$\mstl$$>$$m_{t}+m_{\szi}$ or $m_{b}+m_{\swk}$
and the RB coupling is comparable with the gauge or the top
Yukawa coupling, $\lams/(4\pi)$ $\nle$ $\alpha$, $\alpha_{t}$,
there is a parameter region where
BR($\stl\rightarrow t\szi$) or BR($\stl\rightarrow b\swk$)
competes with BR($\stl\rightarrow ed$).
In Fig.~1 we show $\mstl$ dependence of the
branching ratio of the stop.
Here we take $\mt$=135GeV, $\tan\beta$=2,
$\lambda'_{131}$=0.1, $\tht$=1.0 and
($M_{2}$(GeV), $\mu$(GeV))$=$
($50$, $-100$) for (a) and ($100$, $-50$) for (b).
The output masses for the lightest neutralino and chargino are
($m_{\sz1}$(GeV), $m_{\swl}$(GeV))$=$ ($29$, $71$) for (a)
and ($42$, $71$) for (b).
It is found that if the stop mass $\mstl$ is not too small,
BR($\stl\rightarrow t\szi$) or BR($\stl\rightarrow b\swk$)
dominates over BR($\stl\rightarrow ed$).
In Fig.2 we show the mixing angle $\tht$ dependence of
the branching ratio.
We find that the branching ratio depends also on $\tht$.

In the case of BR($\stl\rightarrow ed$)$\simeq$100\%,
the most promissing process is
the stop $\stl$ production through the $s$-channel in
neutral current (NC) processes (\ref{RBNC}).
We expect its clear signal as a sharp peak in
the Bjorken parameter $x$ distribution and
the peak point corresponds to $x={\mstl^2}/{s}$.
As has been pointed out,
it is well known that the similar peak in the $x$ distribution
could be expected to the leptoquark production at HERA \cite{leptoquark}.
We pointed out that
the stop with the RB couplings could be discriminated from
the most of leptoquarks by its distinctive properties ;
1)the $x$ peak originated from the stop would exist only in
the NC (not exist in the CC) process
because there is no RB stop couplings to the neutrinos,
2)the $e^+$ beams are more favorable than the $e^-$ beams
\cite{proc,full}.
However, one of the leptoquarks ${\widetilde{S}}_{1/2}$
\cite{leptoquark}
with the charge
$Q=-2/3$ would give the same signature as the RB stop.
In fact H1 group at HERA has given the lower mass bound
$\mstl$ $\nge$ 98GeV
on the RB stop from the negative result for the leptoquark
${\widetilde{S}}_{1/2}$ search
at 95\% CL for $\lam$ $=$ 0.3 \cite{H1}.
We should note that this bound is only applicable to
BR($\stl\rightarrow ed$)$\simeq$100\%, i.e.
$\mstl$$<$$m_{t}+m_{\sz1}$ and $m_{b}+m_{\swl}$.
We can see from Figs.1 and 2 that
BR($\stl\rightarrow ed$)$\simeq$100\% will not be realistic
for the heavy stop.
Even for $\lam$ $=$ 0.3 and $\mstl$ $=$ 100GeV
we get BR($\stl\rightarrow ed$)$\simeq$50\%,
where we take ($\mt$, $\tanbe$, $M_2$, $\mu$, $\tht$) $=$
($135$GeV, $2$, $50$GeV, $-100$GeV, $1.0$) for example.
So we should be careful in converting the mass bound on
the leptoquark ${\widetilde{S}}_{1/2}$ into the RB stop.

In the case BR($\stl\rightarrow ed$)$\ll$100\%
the other processes
\begin{equation}
ep\goto t\szi X
\label{top}
\end{equation}
and
\begin{equation}
ep\goto b\swk X
\label{bottom}
\end{equation}
will have viable cross sections to which the stop
contributes from the $s$-channel.
The Feynman diagrams for these processes are depicted in Fig.3.
In these diagrams, we consider also the virtual contributions
of the selectron, sneutrino and $d$-squark with the
same RB couplings $\lambda'_{131}$.
The formulae for differential cross sections are given by
\begin{eqnarray}
&&{\frac{\ddf\hat{\sigma}}{\ddf x\ddf Q^2}}(ep\to t\,\szi X)
= \nonumber \\
&&{\frac{\alpha\lambda^{'2}_{131}}{8\sh^2}}
\Bigg[\left|F_\se\right|^2{\frac{(\uh-\mt^2)(\uh-\mszi^2)}{(\uh-\msel^2)^2}}
+\left|F_\sd\right|^2{\frac{(\th-\mt^2)(\th-\mszi^2)}{(\th-\msdr^2)^2}}
\nonumber \\
&&+{\frac{\cos^2\tht \sh}{\stg}}
\Big(\left(\left|F_L\right|^2+\left|F_R\right|^2\right)
\left(\sh-\mt^2-\mszi^2\right)-4\mt\mszi{\rm Re}\left(F_RF_L^*\right)
\Big)
\nonumber \\
&&+2{\rm Re}\left(F_\se F_\sd^*\right){\frac{\th\uh-\mt^2\mszi^2}
{(\uh-\msel^2)(\th-\msdr^2)}}
\nonumber \\
&&+{\frac{2\cos\tht\sh(\sh-\mstl^2)}{(\stg)(\uh-\msel^2)}}
{\rm Re}\left(F_\se^*(F_R\uh+F_L\mt\mszi)\right)
\nonumber \\
&&+{\frac{2\cos\tht\sh(\sh-\mstl^2)}{(\stg)(\th-\msdr^2)}}
{\rm Re}\left(F_\sd^*(F_R\th+F_L\mt\mszi)\right) \Bigg],
\end{eqnarray}
where $\sh=xs$, $\th=-Q^2$ and
\begin{eqnarray}
F_\se&\equiv& e_eN'_{i1}-{\frac{1/2+e_e\sin^2\tew}{\cos\tew\sin\tew}}
N'_{i2}, \\
F_\sd&\equiv& e_dN'_{i1}-{e_d\tan\tew}N'_{i2}.
\end{eqnarray}
\begin{eqnarray}
&&{\frac{\ddf\hat{\sigma}}{\ddf x\ddf Q^2}}(ep\to b\,\swk X)
= \nonumber \\
&&{\frac{\alpha\lambda^{'2}_{131}}{16\sin^2\tew\sh^2}}
\Bigg[\left|V_{11}\right|^2{\frac{(\uh-\mb^2)(\uh-\mswk^2)}{(\uh-\msn^2)^2}}
\nonumber \\
&&+{\frac{\cos^2\tht \sh}{\stg}}
\Big(\left(\left|G_L\right|^2+\left|G_R\right|^2\right)
\left(\sh-\mb^2-\mswk^2\right)-4\mb\mswk{\rm Re}\left(G_RG_L^*\right)
\Big)
\nonumber \\
&&-{\frac{2\cos\tht\sh(\sh-\mstl^2)}{(\stg)(\uh-\msn^2)}}
{\rm Re}\left(V_{11}^*(G_R\uh+G_L\mb\mswk)\right) \Bigg].
\end{eqnarray}
Figure 4 shows the stop mass dependence of the total cross sections
for $e^{-}p$ and $e^{+}p$ collisions.
It is found that to get larger cross section
the $e^+$ beam is more efficient than the $e^-$ one.
This can easily be understood from the structure of the coupling.
While the $e^-$ collides only with sea $\bar{d}$-quarks in
the proton, the $e^+$ collides with valence $d$-quarks.
The difference of structure functions of the proton is naturally
reflected in the cross sections.
{}From Fig.4 we expect the detectable cross sections $\nge$ 0.1pb for
heavy stop with the mass $\mstl\nle$250GeV if we use the $e^+$ beam.
Only the process $ep\goto b\swl X$ would be detectable
for $\mstl\nle$170GeV so long as we use the $e^-$ beam.

Next we should discuss the signature of these processes.
Note that the LSP, the lightest neutralino $\sz1$, will decay
into $R$-even particles via the RB couplings \cite{susana}.
In our model with only non-zero RB coupling $\lambda'_{131}$,
$\sz1$ decays into $bd\nu$ (see Fig.~5).
Then the typical decay chains are
\begin{equation}
ep\goto t\sz1 X \goto (bW)(bd\nu)X \goto (b(\ell\nu))(bd\nu)X
\label{tsZ}
\end{equation}
and
\begin{equation}
ep\goto b\swl X \goto b(\ell\nu\sz1)X \goto b(\ell\nu(bd\nu))X.
\label{bsW}
\end{equation}
In both processes one of typical signatures would be
2 $b$-jets $+$ jet $+$ $\ell$ $+$ $\misP$.
In Fig.6 we show the Monte Carlo events for the
transverse momentum distribution of the scattered muon from
the process (\ref{bsW})
under the condition of the integrated luminosity $L=300$pb$^{-1}$
for $e^{-}p$ collisions.
For simplicity,
BR($\swl\goto \mu\nu\sz1$) is assumed to be $1/9$ \cite{Tata}.
Here we depict also the possible background muon events.
They come from both charged current (CC) processes
$e^-p\goto \nu q X$ and $W$-gluon fusion (WGF) processes
$e^-p\goto \nu s\overline{c} X$, $\nu b\overline{c} X$,
where use has been made of the generators LEPTO\cite{LEPTO}
and AROMA\cite{AROMA} with JETSET\cite{JETSET}, respectively.
We find that the lower $p_T$ cut for the scattered muon could be useful
to distinguish the process (\ref{bottom}) from the backgrounds.
Since the multi-jet events accompanied by
the high $p_T$ muon are a distinctive signature
for the RB stop production,
the stop could be discriminated from the leptoquark
${\widetilde{S}}_{1/2}$ with the charge $Q=-2/3$.

  Now we summarize our results obtained here.
We have investigated various production processes
of the stop at HERA energies in the framework of the
MSSM with the RB coupling of the stop.
If the stop is light enough
$\mstl$$<$$m_{t}+m_{\sz1}$, $m_{b}+m_{\swl}$,
the stop produced via RB interactions shows a sharp peak in the
$x$ distribution of neutral current processes
due to its $s$-channel resonance.
However,  it is difficult to discriminate
the stop from one of the leptoquarks ${\widetilde{S}}_{1/2}$.
On the other hand, the other processes
$ep\goto t\sz1 X$ and $ep\goto b\swl X$
will have viable cross sections to which the stop
contributes from the $s$-channel for $\mstl$ $\nge$ 100GeV.
In both processes one of the typical signatures would be
2 $b$-jets $+$ jet $+$ $\ell$ $+$ $\misP$
owing to the LSP decay.
One of the detectable signals of these processes is
characterized by high $p_{T}$ spectra of muons
which are rather different from those of the background processes.
Since this is a distinctive signature
the stop could be discriminated from the leptoquark
${\widetilde{S}}_{1/2}$.

\vglue 0.5cm


\vglue 0.4cm


\vfill\eject

\baselineskip = 18pt plus 1pt
\noindent{\Large{\bf Figure Captions}} \\
\medskip
\nobreak
{\bf Figure 1:} \ \
$\mstl$ dependence of branching ratio of stop.
We take $\mt$=135GeV, $\tan\beta$=2, $\tht$=1.0,
$\lambda'_{131}$=0.1 and
($M_{2}$(GeV), $\mu$(GeV))$=$
($50$, $-100$) for (a) and ($100$, $-50$) for (b).
\\
\medskip
{\bf Figure 2:} \ \
$\tht$ dependence of branching ratio of stop.
We take $\mt$=135GeV, $\tan\beta$=2,
$\lambda'_{131}$=0.1,
$M_{2}$=50GeV, $\mu$=$-300$GeV and $\mstl$=200GeV.
 \\
\medskip
{\bf Figure 3:} \ \
Feynman diagrams for sub-processes
$eq \goto t\szi$ and $eq \goto b\swk$.
 \\
\medskip
{\bf Figure 4:} \ \
Stop mass dependence of total cross section.
We take $\mt$=135GeV, $\tht$=1.0, $\tan\beta$=2,
$\lambda'_{131}$=0.1 $M_{2}$=100GeV, $\msl$=200GeV, $\msq$=300GeV
and $\mu$=$-50$GeV.
Solid, sort-dashed, dotted and dashed lines correspond to
$e^{-}p \goto b\swl^{-}X$, $e^{+}p \goto \bar{b}\swl^{+}X$,
$e^{-}p \goto \bar{t}\sz1X$ and $e^{+}p \goto {t}\sz1X$, respectively.
 \\
\medskip
{\bf Figure 5:} \ \
Feynman diagrams for LSP decay.
\label{fig5}
 \\
\medskip
{\bf Figure 6:} \ \
Monte Carlo events for transverse momentum distribution of
scattered muon from $e^{-}p\goto b\swl X$ (solid lines)
together with backgrounds CC and WGF processes (short-dashed line).
We take $\msl$=200GeV, $\msq$=300GeV,
$\mt$=135GeV, $\tht$=1.0, $\tan\beta$=2,
$\lambda'_{131}$=0.1 $M_{2}$=100GeV, $\mu$=$-50$GeV and
integrated luminosity $L=300$pb$^{-1}$.
 \\
\vfill\eject

\end{document}